\documentclass[5p,compress]{elsarticle}

\date{January 9, 2022}
\journal{Physics Letters B}
\usepackage{newtxtext}
\usepackage[varvw,bigdelims]{newtxmath}
\usepackage[colorlinks=true,allcolors=blue]{hyperref}
\usepackage{lineno}
\modulolinenumbers[5]

\usepackage{amsmath}
\newcommand{\nuc}[2][]{{}^{#1}\mathrm{#2}}

\begin{document}

\begin{frontmatter}

\title{Phase-space consideration on barrier transmission in a time-dependent variational approach with superposed wave packets}
\author{Akira Ono}
\ead{ono@nucl.phys.tohoku.ac.jp}
\address{Department of Physics, Tohoku University, Sendai 980-8578, Japan}

\begin{abstract}
  A known limitation of time-dependent mean-field approaches is a lack of quantum tunneling for collective motions such as in sub-barrier fusion reactions. As a first step toward a solution, a time-dependent model is considered using a superposition of Gaussian wave packets, to describe the relative motion between two colliding nuclei, which may be simplified to a problem for one particle in one dimension. In this article, how the model describes the potential-barrier transmission is investigated by paying attention to the time evolution of the phase space distribution, which in particular reveals that the behavior of the free propagation of the incoming state is not trivial, depending on the number of superposed wave packets. Passage over the barrier can occur due to the high-momentum components in the incoming state corresponding to energies above the barrier height, which is, however, of classical nature and needs to be distinguished from the true quantum tunneling. Although a transmitted wave packet in some case may end up with an energy lower than the barrier, a difficulty is noticed in guaranteeing the energy conservation when the energies of different exit channels, e.g.\ of transmission and reflection, are individually measured. To overcome these issues for a description of quantum tunneling is still a challenging problem. This article mainly treats the same system with the same model as in the paper \href{https://doi.org/10.1016/j.physletb.2020.135693}{Phys.\ Lett.\ B 808 (2020) 135693}, \href{https://arxiv.org/abs/2006.06944v1}{arXiv:2006.06944v1} by N.~Hasegawa, K.~Hagino and Y.~Tanimura. However, the conclusion of the present work disagrees with their quick conclusion that quantum tunneling was simulated by the model. Comments are made on this.
\end{abstract}


\end{frontmatter}

%\linenumbers
\section{Introduction}
It is a long-standing problem to describe many-particle quantum tunneling in time-dependent mean-field approaches \cite{hasegawa2020,negele1982,simenel2012,nakatsukasa2016}, such as for sub-barrier fusion of colliding nuclei \cite{hagino2012}. To predict emergence of multiple reaction channels, e.g.\ of fusion and scattering, is a general desire in time-dependent approaches for many-body systems. In some cases, stochastic extension of a model can allow channels to emerge as a consequence of time evolution, such as in transport models for heavy-ion collisions \cite{ono2019ppnp, colonna2020ppnp, xu2019ppnp, ono2004ppnp, chomaz2004, ono1996amdv, ono2002, colonna1994}. Tunneling further requires true quantum description for the translational motion of a nucleus, which is often not straightforward because the center-of-mass wave function is enforced to be localized in space, like a wave packet, when the nucleus is described in a mean-field model.

In this situation, a possible idea to go beyond is to use a coherent superposition of Slater determinants
\begin{equation}
|\Psi\rangle=\sum_{a}f_a|\Phi(z_a)\rangle,
\end{equation}
where each Slater determinant $|\Phi(z_a)\rangle$ is specified by a set of parameters $z_a$. It was suggested by Ref.~\cite{hasegawa2020} to determine the time dependence of these parameter sets $\{z_a(t)\}$ and the coefficients $\{f_a(t)\}$ by the time-dependent variational principle, to do a simulation starting with a given initial condition. This formalism was called a time-dependent generator coordinate method (TDGCM). Among the models which may be called by the same name \cite{reinhard1983, orestes2007, regnier2019, zdeb2017, regnier2016, bernard2011, goutte2005, tao2017}, the model used in Ref.~\cite{hasegawa2020} is unique in that both $\{z_a(t)\}$ and $\{f_a(t)\}$ are simultaneously determined.

For the Slater determinants $|\Phi(z_a)\rangle$ in TDGCM, we use in this article the wave functions of antisymmetrized molecular dynamics (AMD) \cite{ono1992,kanada2012}, i.e., Slater determinants of Gaussian wave packets of a fixed width. One of the benefits of using an AMD wave function for $|\Phi(z_a)\rangle$ is that the center-of-mass motions of composite particles are separated from their internal degrees of freedom. Therefore, e.g.\ for a collision of $\nuc[4]{He}+\nuc[4]{He}$, the many-body state can be rewritten as
\begin{equation}
\Psi=\mathcal{A}\bigl[\psi(x)\Phi_\alpha(\xi_1)\Phi_\alpha(\xi_2)\bigr]\Phi_{\text{c.m.}}(X_{\text{c.m.}}),
\label{eq:clustergcm}
\end{equation}
when the internal excitations of each nucleus are ignored. Here the center-of-mass wave function $\Phi(X_{\text{c.m.}})$ and the internal wave functions of individual $\nuc[4]{He}$ nuclei, $\Phi_\alpha(\xi_1)$ and $\Phi_\alpha(\xi_2)$, are all written with Gaussian functions in AMD. Then, in this simplified example, only the time evolution of the wave function for the relative coordinate $x$ between the two nuclei
\begin{equation}
\psi(x,t)=\sum_af_a(t)\, e^{-\nu_r\bigl(x-\frac{z_a(t)}{\sqrt{\nu_r}}\bigr)^2}
\label{eq:relativewf}
\end{equation}
needs to be considered in the many-body state of Eq.~\eqref{eq:clustergcm}. The Gaussian centers $z_a(t)$ take complex values. It should be noted that $z_a(t)$ do not always have direct physical meaning when non-orthogonal wave packets are coherently superposed. The width parameter for the relative motion is $\nu_r=2\nu$ in this example, where $\nu$ is that of single-particle wave packets.

The aim of this paper is to investigate how the TDGCM model predicts barrier transmission in a simplified problem for one degree of freedom, still keeping the source of the issue in many-body problems that the state usually has to be described by localized wave packets. To allow intuitive understanding, in Sec.~\ref{sec:incoming}, we pay attention to the time evolution of the phase-space distribution function which directly represents the information on the wave function $\psi(x,t)$ of Eq.~\eqref{eq:relativewf}. In particular, already in the incoming state, we will realize important roles played by the momentum distribution. Considerations are made in Sec.~\ref{sec:passage} based on the phase space, to clarify how the barrier transmission occurs in TDGCM, and what problems have to be overcome to extract the information on quantum tunneling, if any, from TDGCM simulations.

This study mainly considers the same setup of the reaction system with the same TDGCM model as the study of Ref.~\cite{hasegawa2020}. In fact, numerical results published there are partly used here in Secs.~\ref{sec:incoming} and \ref{sec:passage}. However, the consequence of the present study disagrees with the quick conclusion made by Ref.~\cite{hasegawa2020} that TDGCM could simulate quantum tunneling. Comments are made on this in Sec.~\ref{sec:comments}.

\section{Problem setup}
Our consideration here is limited to the cases which are reduced to problems of one particle in one dimension, including the case of $\nuc[4]{He}+\nuc[4]{He}$ in one dimension. We also assume that the antisymmetrization $\mathcal{A}$ in Eq.~\eqref{eq:clustergcm} is not essential. We will use the terminology for one-particle problems below for simplicity. The translation to the case of the relative motion of two nuclei is straightforward.

For the potential barrier $V(x)$, we should allow various possibilities of the barrier height and width for the purpose to test the model. However, also for the simplicity of terminology, we assume that $V(x)$ is sufficiently smooth compared to the spatial width of the wave packets, so we do not distinguish $V(x)$ and its expectation value for a wave packet centered at $x$.

The model uses compact wave packets for the expansion in Eq.~\eqref{eq:relativewf}. Gaussian wave packets form an overcomplete non-orthogonal basis, so $\psi(x,t)$ of Eq.~\eqref{eq:relativewf} can express the exact solution of the Schr\"odinger equation if a sufficiently large number of wave packets can be superposed. On the other hand, the study in Ref.~\cite{hasegawa2020} was done when two wave packets are used for expansion as
\begin{equation}
\psi(x,t)=f_1(t)\,e^{-\nu_r\bigl(x-\frac{z_1(t)}{\sqrt{\nu_r}}\bigr)^2}
+f_2(t)\,e^{-\nu_r\bigl(x-\frac{z_2(t)}{\sqrt{\nu_r}}\bigr)^2}.
\label{eq:gcmwf}
\end{equation}
A result with 10 wave packets was also reported briefly in Ref.~\cite{hasegawa2020}. It is not clear whether one should ideally superpose as many wave packets as possible until the result converges, or one should choose a suitable number of wave packets for the best prediction of quantum tunneling. We will consider both cases of philosophy of the model.

The position and momentum centers of each Gaussian wave packet, denoted here by $x_a(t)$ and $p_a(t)$, are related to the parameter $z_a(t)$ by
\begin{equation}
z_a(t)=\sqrt{\nu_r}\,x_a(t) + i\frac{p_a(t)}{2\hbar\sqrt{\nu_r}}.
\end{equation}
We choose the same width parameter $\nu_r= 1\ \text{fm}^{-2}$ as Ref.~\cite{hasegawa2020}. As an example of the initial condition, we may consider the same condition as in Ref.~\cite{hasegawa2020} by taking
\begin{equation}
x_1-x_2=0.1\ \text{fm}\quad\mbox{and}\quad
p_1-p_2=1.21\ \text{MeV}/c
% \begin{split}
% \bigl(x_1, p_1\bigr)
% &=\bigl(-30.0\ \text{fm},\ 20.59\ \text{MeV}/c\bigr)\\
% \bigl(x_2, p_2\bigr)
% &=\bigl(-30.1\ \text{fm},\ 19.37\ \text{MeV}/c\bigr)
% \end{split}
\label{eq:inicond}
\end{equation}
for the two wave packets at $t=0$. The coefficients are taken to be $f_1=f_2$ at $t=0$.

\begin{figure}\centering
\includegraphics[width=0.35\textwidth]{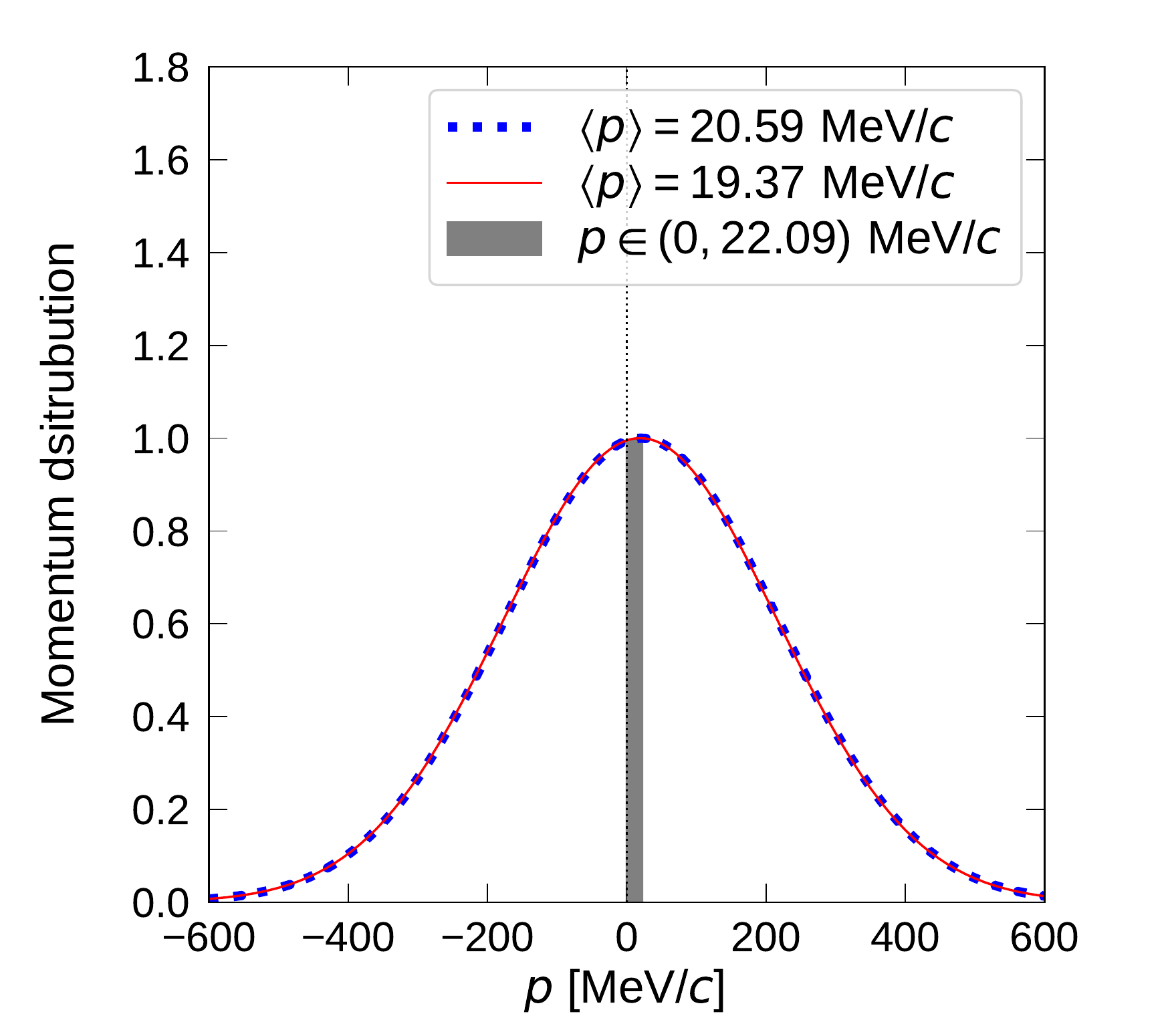}
\caption{\label{fig:momdst}
The momentum distributions for the two Gaussian wave packets which were used in the initial state in the calculation of Ref.~\cite{hasegawa2020}. The gray area indicates the region that is relevant to quantum tunneling ($E<0.13$ MeV and $p>0$).
}
\end{figure}

In Ref.~\cite{hasegawa2020}, the initial momentum was chosen as $\langle p\rangle=\frac12(p_1+p_2)=19.98$ MeV/$c$, which is much smaller than the momentum width, $\Delta p=\hbar\sqrt{\nu_r}=197$ MeV/$c$, associated with the Gaussian wave packet. In Fig.~\ref{fig:momdst}, the momentum distributions for the two individual components in Eq.~(\ref{eq:gcmwf}) at $t=0$ are shown by the blue dotted line (for the first term) and by the red solid line (for the second term). The momentum centers $p_1$ and $p_2$ are practically identical. Since the two wave packets are superposed in phase in Eq.~(\ref{eq:gcmwf}), the distribution for the total wave function $\psi(x,t=0)$ also has the same shape.  An important question to be addressed later is whether this broad momentum width plays any role for the time evolution of $\psi(x,t)$. For the choice of the potential $V(x)$ in Ref.~\cite{hasegawa2020}, the gray area indicates the region of $0<p<\sqrt{2\mu V_{\text{B}}}$, where $V_{\text{B}}=0.13$ MeV is the barrier height. The mass is $\mu=1876$ MeV/$c^2$. This narrow region is the part relevant to quantum tunneling, where the potential barrier cannot be passed over in classical dynamics.

\section{Dual view on energy and momentum \label{sec:views}}

To avoid confusions, let us clearly distinguish two possible viewpoints on the interpretation of the state described by the wave function $\psi(x,t)$, or equivalently by the parameters such as $x_a(t)$, $p_a(t)$ and $f_a(t)$.  Of course, in the usual sense of quantum mechanics, we can extract any physical information from $\psi(x,t)$ using the standard recipes. Let us call this way ``View A'' in this article. In the other viewpoint, which we call ``View B'' here, one ignores the momentum width of individual wave packets, so that a wave packet is regarded as having a definite momentum $p_a$. More generally, the state represented by $\psi(x,t)$ might be assumed to have a definite momentum calculated as $\langle\psi|p|\psi\rangle/\langle\psi|\psi\rangle$. A clear difference between the two views can be identified by considering the value of kinetic energy. In View A, the expectation value is $\langle p^2/2\mu\rangle=p_a^2/2\mu+T_0$, where the constant term \begin{equation}
T_0=\hbar^2\nu_r/2\mu
\end{equation}
originates from the momentum width of the wave packet. On the other hand, in View B, the wave packet has a definite kinetic energy $E_a=p_a^2/2\mu$. The difference between the two views is not negligible; $T_0=10.38$ MeV in our example.

When nuclear reactions are simulated, View B has been taken widely in time-dependent approaches, such as TDHF and transport models including AMD, for the center-of-mass motions of nuclei. This view is convenient because the reaction with a fixed beam energy can be simulated by choosing the initial value $p_a=\sqrt{2\mu E_{\text{beam}}}$ and also because an emitted nucleus can be regarded as having a definite energy $E_a=p_a^2/2\mu$. Although View B violates rules of quantum mechanics (e.g.\ Heisenberg's uncertainty principle), it works in many cases without causing serious inconsistencies, mainly because $T_0$ is just a constant term in the Hamiltonian, so it affects neither the equation of motion nor the energy conservation. However, to the best of the author's knowledge, the models that take View B have not been extended successfully to treat coherent superposition of wave packets, though stochastic branching of wave packets or many-body states is typically considered in transport models \cite{ono2019ppnp, colonna2020ppnp, xu2019ppnp, ono2004ppnp, chomaz2004, ono1996amdv, ono2002, colonna1994}. It is therefore an important question whether View B can be taken in the TDGCM model, which will be addressed later in detail.

We are relatively safe in View A, though a drawback is that the initial state cannot be prepared as a momentum eigenstate.

The lack of quantum tunneling in the models with a single wave packet (i.e.\ a single Slater determinant) can be understood in either view. In View B, the classical Hamiltonian without the $T_0$ term can simply explain that the barrier cannot be passed over when $p_a<\sqrt{2\mu V_{\text{B}}}$ ($a=1$). View A should be based on the initial momentum distribution $f(p)$ associated with the wave packet. When $p_a<\sqrt{2\mu V_{\text{B}}}$, a tail of $f(p)$ may be still extending to the high-momentum region ($p>\sqrt{2\mu V_{\text{B}}}$) where barrier passage would be possible. However, the probability in $f(p)$ is larger in the part of $p<\sqrt{2\mu V_{\text{B}}}$ where barrier passage is less probable even if quantum tunneling is allowed. Since the model does not allow the wave packet to split into two parts, the dynamics can only choose the major branch so the wave packet will be totally reflected. On the other hand, when $p_a>\sqrt{2\mu V_{\text{B}}}$, the wave packet will be totally transmitted for the same but inverted reason. In this case, some tail component of $f(p)$ may still exist in the low-momentum region, and therefore one could in principle argue that transmission always occurred even when there was some probability of $p^2/2\mu<V_{\text{B}}$. However, this is customarily not called quantum tunneling.

\section{\label{sec:incoming} Incoming state}

\begin{figure*}
\includegraphics[width=\textwidth]{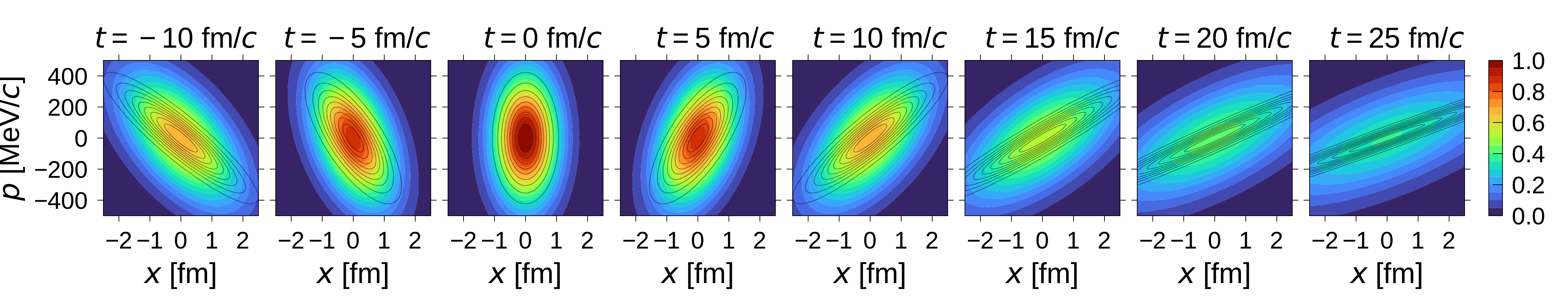}
\caption{\label{fig:freef}
Wigner and Husimi functions for a free particle motion, at different times from $t=-10$ fm/$c$ to 25 fm/$c$. The Wigner function is shown by the contour lines drawn for $f=0.2, 0.4, 0.6,\ldots$, while the Husimi function is represented by the color scale. The initial condition for the Wigner function was chosen as $f_{\text{W}}(x,p,t=0)=2\exp(-2\nu_rx^2-p^2/2\hbar^2\nu_r)$ which is very similar to the Wigner function of the initial state $\psi(x,t=0)$ of the TDGCM calculation in Ref.~\cite{hasegawa2020}. 
}
\end{figure*}

A freely propagating state is used at the initial stage of the simulation. Free propagation is trivial in many models, but it may not be so in a newly invented model, so we should carefully consider it in TDGCM. In fact, we will realize that the incoming state in the TDGCM simulation has special features.

\subsection{Exact solution and superposition of many wave packets}

The well known exact solution for a free motion is displayed in Fig.~\ref{fig:freef}, in which the Wigner function $f_{\text{W}}(x,p,t)$ \cite{wigner1932} is shown from $t=-10$ to $25$ fm/$c$, by solid contour lines. The figure was drawn in the boosted frame in which $\langle p\rangle=0$ and $\langle x\rangle=0$. Importantly, the correlation between $x$ and $p$ develops strongly as the time progresses, which indicates that different momentum components are resolved by the time evolution, resulting in different spatial positions. As a reference for comparison, we will use later the Husimi function \cite{husimi1940} defined by
\begin{equation}
f_{\text{H}}(x,p,t)=\iint\frac{dx'dp'}{2\pi\hbar}e^{-\frac{(x-x')^2}{2\Delta x^2}-\frac{(p-p')^2}{2\Delta p^2}}f_{\text{W}}(x',p',t)
\label{eq:husimiwigner}
\end{equation}
with $\Delta x=1/(2\sqrt{\nu_r})$ and $\Delta p=\hbar\sqrt{\nu_r}$. It is shown in Fig.~\ref{fig:freef} by the color scale.

The equation for the time evolution of $f_{\text{W}}(x,p,t)$, derived from the Schr\"odinger equation, can be viewed as an extension of the Liouville equation with quantum terms in general (see e.g.\ Refs.~\cite{takahashi1989, hwlee1995, tannor2007}). However, in the particular cases of the potential of the form $V(x)=ax^2+bx+c$ including the case of free propagation, the equation is exactly identical to the classical Liouville equation. Therefore, the evolution of the distribution function can be easily understood based on the classical trajectories in the phase space.

If a sufficiently large number of wave packets are superposed as in Eq.~\eqref{eq:relativewf} in a suitable way, the free propagation can be described precisely by TDGCM in principle. For example, wave packet centers, all of which are prepared almost at the same point in the phase space, will move to reproduce the exact solution e.g.~by aligning along the principal axis of the deformed distribution, with suitable changes of weight factors.

\begin{figure}
\centering
\includegraphics[width=0.5\textwidth]{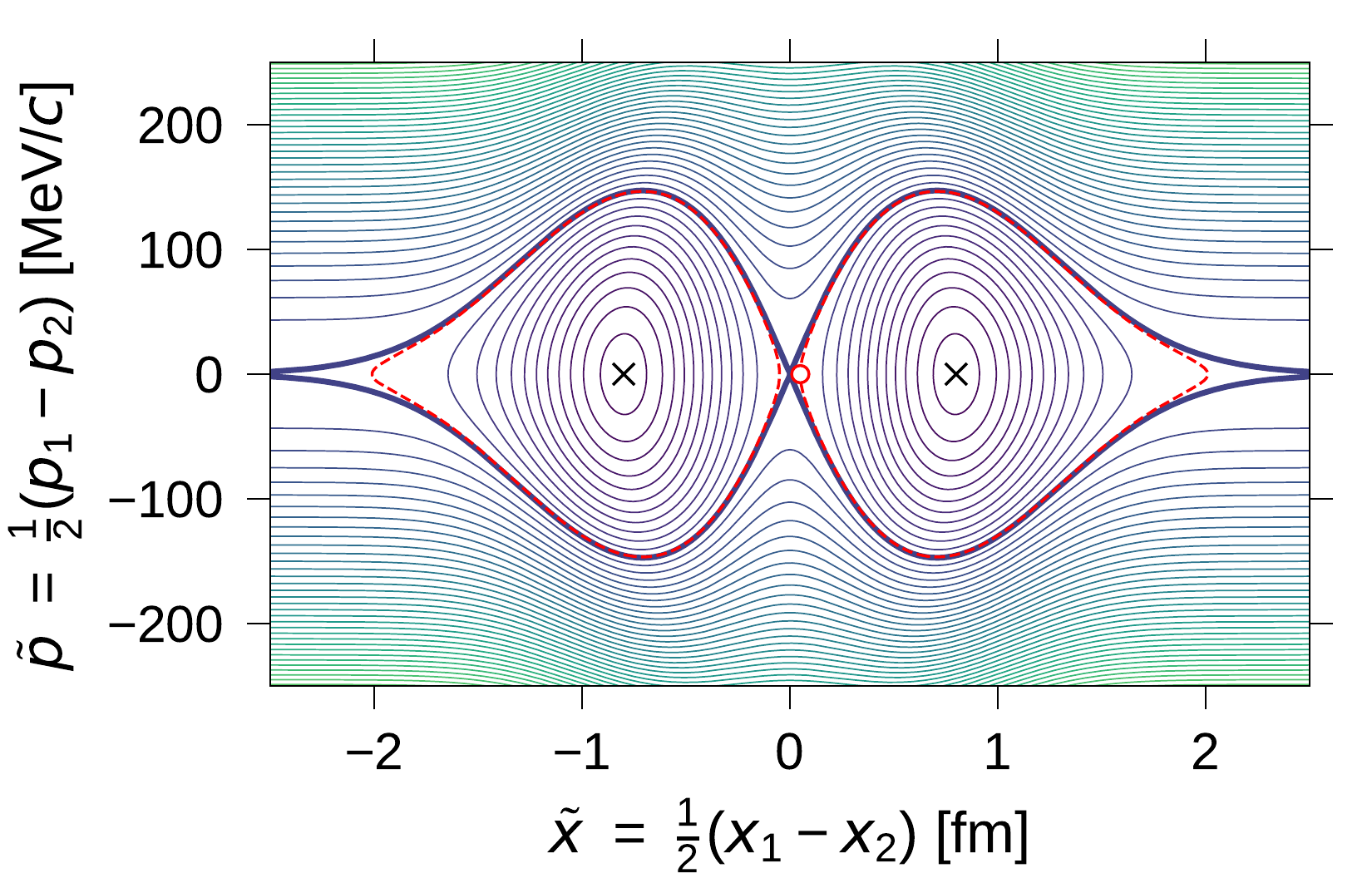}
\caption{\label{fig:ekincontour} Contour plot for the energy $E$ of a free particle when the wave function is approximated by a superposition of two Gaussian wave packets, centered at $(\tilde{x},\tilde{p})$ and $(-\tilde{x},-\tilde{p})$, with equal coefficients that are in phase.  Contour lines are drawn for $E=\hbar^2\nu_r/2\mu + k\times 0.5\ \text{MeV}$ for integers $k$. The thick contour line is for $E=\hbar^2\nu_r/2\mu=10.38$ MeV.  The energy takes the minimum $E=4.60$ MeV at $(\tilde{x}, \tilde{p})=(\pm0.80\ \text{fm}, 0)$, shown by the crosses. The open circle indicates the initial condition chosen in Ref.~\cite{hasegawa2020}, $(\tilde{x}, \tilde{p})=(0.05\ \text{fm},\ 0.61\ \text{MeV}/c)$. The dashed contour line is for $E= 10.33$ MeV corresponding to this initial condition. The width parameter of the Gaussian wave packets and the mass are chosen to be $\nu_r=1\ \text{fm}^{-2}$ and $\mu=1876$ MeV.}
\end{figure}

\begin{figure*}
\includegraphics[width=\textwidth]{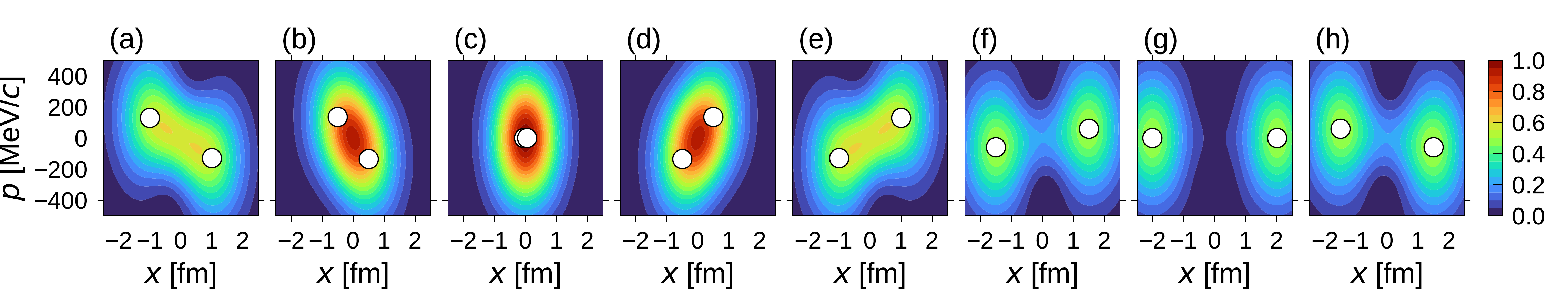}
\caption{\label{fig:husimif}
Husimi function for the TDGCM state of Eq.~(\ref{eq:gcmwf}) in a boosted frame. Each panel represents the Husimi function by the color scale, for the state specified by the parameters $(\tilde{x}, \tilde{p})$, whose values are chosen on the dashed contour line in Fig.~\ref{fig:ekincontour}. The open circles represent the locations of the wave packet centers, $\pm(\tilde{x},\tilde{p})$.
%
% (a): $(1.00,    -129.53)$,
% (b): $(0.50,    -135.53)$,
% (c): $(0.05,       0.00)$,
% (d): $(0.50,     135.53)$,
% (e): $(1.00,     129.53)$,
% (f): $(1.50,      59.37)$,
% (g): $(2.01,       0.00)$,
% (h): $(1.50,     -59.37)$.
%
}
\end{figure*}

\subsection{Superposition of two wave packets}

Next, let us investigate the free propagation when only two wave packets are superposed in TDGCM. We can easily know the trajectories of the wave packet centers from the energy contours, as shown in Fig.~\ref{fig:ekincontour}. The energy expectation value is
\begin{align}
E=\langle H\rangle
&=T_0
+\frac{1}{N}\sum_a\sum_b \frac{p_{ab}^2}{2\mu} f_a^*f_bN_{ab}
\label{eq:ekin}
\\
N&=\sum_a\sum_b f_a^*f_bN_{ab}
\end{align}
for a TDGCM state, with $p_{ab}=i\hbar\sqrt{\nu_r}(z_a^*-z_b)$ and the overlap matrix $N_{ab}=e^{-\frac{1}{2}(z_a^*-z_b)^2}$.  For a free motion, the condition of $f_1=f_2$ is preserved and the wave packet centers can be expressed as $(x_1, p_1)=(\tilde{x}, \tilde{p})$ and $(x_2, p_2)=(-\tilde{x}, -\tilde{p})$ at any time $t$ in the boosted frame.  The thick contour line in Fig.~\ref{fig:ekincontour} is for $E=T_0$. With this line as a separatrix, there are two kinds of trajectories of $(\tilde{x},\tilde{p})$. If $|\tilde{p}|$ is sufficiently large, $|\tilde{x}|$ will increase, resulting eventually in non-interfering two wave packets. On the other hand, in the region surrounded by the separatrix, cyclic trajectories exist. They move around the energy minima shown by the crosses in Fig.~\ref{fig:ekincontour}. The energy at the minimum configuration is $E=4.60$ MeV, which is significantly lower than $T_0=10.38$ MeV, suggesting that two wave packets can form a rigid pair by effectively constructing a single wave packet with a narrow momentum width and a broad spatial width.

The point of $(\tilde{x},\tilde{p})$ of the initial condition Eq.~(\ref{eq:inicond}) is indicated in Fig.~\ref{fig:ekincontour} by the open circle. It was chosen inside the domain of cyclic trajectories, but it is very close to the separatrix. Under this initial condition, the point of $(\tilde{x},\tilde{p})$ should move cyclically on an energy contour shown by the red dashed line in Fig.~\ref{fig:ekincontour}. This is consistent with the trajectories in Fig.~2 of Ref.~\cite{hasegawa2020}. Here we are only looking at the wave packet centers, as in View B (see Sec.~\ref{sec:views}). The physical meaning of a cyclic motion for free propagation seems difficult to understand in this view.

Let us then take View A by looking directly at the information in the TDGCM wave function $\psi(x,t)$ of Eq.~(\ref{eq:gcmwf}). Eight points are chosen on the red dashed contour in Fig.~\ref{fig:ekincontour}, and the Husimi function $f_{\text{H}}(x,p)$ is calculated for each point and displayed in Fig.~\ref{fig:husimif}. In each panel, the location of the superposed two wave packets, $\pm(\tilde{x},\tilde{p})$, are shown by open circles. The Husimi function is defined by Eq.~(\ref{eq:husimiwigner}) with the Winger function
\begin{align}
f_{\text{W}}(x,p)
&=\int ds\, e^{-ips/\hbar}\,\psi^*(x-\tfrac12s)\,\psi(x+\tfrac12s)\\
&=\frac{1}{N}\sum_a\sum_b 2\,e^{-2\nu_r(x-x_{ab})^2-(p-p_{ab})^2/2\hbar^2\nu_r}f_a^*f_b N_{ab},
\end{align}
with $x_{ab}=(z_a^*+z_b^*)/2\sqrt{\nu_r}$ and $p_{ab}=i\hbar\sqrt{\nu_r}(z_a^*-z_b)$. The Wigner function is, however, not displayed in the figure; it sometimes shows violent oscillatory behavior and it takes negative values in some regions. 

The panels in Fig.~\ref{fig:husimif} have been arranged so that they are similar to those for the exact solution in Fig.~\ref{fig:freef}, at least for the left five panels. In fact, the evolution in panels from (b) to (d), is quite similar to the exact evolution from $t=-5$ to $5$ fm/$c$. The states in (a) and (e) may also be acceptable approximations at $t\approx -10$ and $10$ fm/$c$. In this time interval, the evolution is consistent with the correct development of the $p$--$x$ correlation. However, after the state of (e), the TDGCM solution largely deviates from the exact solution. The distribution is unphysically separated into two parts in (f), (g) and (h). The shape of each part is almost frozen, and the evolution of the $p$--$x$ correlation is now pathological. In panel (f), for example, the front part ($x>0$) of the distribution still has a weak tendency to have positive momentum $\langle p\rangle$ on average, but this part has to unphysically lose $\langle p\rangle$, in order to move to the forward direction, stalling at the state of (g). The TDGCM solution further evolves to (h) and then to (a), closing a cycle of the trajectory. In Fig.~2 of Ref.~\cite{hasegawa2020}, the period of a cycle is found to be about 150 fm/$c$, which is much longer than the time from panel (a) to (e) in Fig.~\ref{fig:husimif}. Namely, the pathological motion, from (e) through (g) to (a), is very slow.

Coming back to View B, let us look at the momentum $p_a$ and the energy $E_a=p_a^2/2\mu$ ($a=1$ or $2$) for this free propagation. The consequence of the model is that both $p_a$ and $E_a$ are changing in time, which sounds quite unrealistic as a free motion. The amount of the change is of the order of 100 MeV/$c$ in $p_a$, which is similar to the momentum width for the wave function. This implies that the origin of the change of $p_a$ and $E_a$ is the momentum width, or the evolution of the $p$--$x$ correlation induced by it. The effect of coherent superposition is not easy to understand in View B.

The state studied here is used as the incoming state of a TDGCM simulation. If one takes View B, it is at least inconvenient that the energy is changing so much before the potential plays any role. In View A, the state has a momentum distribution $f(p)$ which has to be taken care of in some way, but a good feature is that the time dependence of $f(p)$ during the free propagation is small because the mean and the variance of the distribution are exactly conserved by the model. However, the evolution of the $p$--$x$ correlation is pathological, as in Fig.~\ref{fig:husimif}.

\section{\label{sec:passage} Considerations on barrier passage}

The incident particle is described as in the previous section by a time-dependent state in TDGCM, and it now arrives at the potential barrier. We have to carefully consider how the behavior of the incoming state may affect the barrier passage.

\subsection{Superposition of many wave packets}

First, let us consider the case in which a large number of wave packets are superposed. In View A, the evolution in the incoming stage will be similar to the exact free propagation in Fig.~\ref{fig:freef}. Note that the initial state has some momentum distribution $f(p)$. The potential barrier is first encountered by the leading front part of the phase space distribution which has a relatively high momentum as a consequence of the developed $p$--$x$ correlation. This leading part can pass over the barrier if it has energy higher than the barrier $V_{\text{B}}$, and the part arriving later will tend to be reflected if the energy is lower than $V_{\text{B}}$. Similar consideration is possible in View B for the passage and reflection of individual wave packets, which have been aligned to represent the developed $p$--$x$ correlation during the free propagation.

When a moderate number (e.g.\ 10) of wave packets are superposed, the situation may be similar to the above consideration (see e.g.\ the case with 10 wave packets briefly reported in Ref.~\cite{hasegawa2020}, and that in Ref.~\cite{hasegawa2021jps}). However, 10 wave packets are probably not enough to precisely express the free propagation for a long time (for $3000$ fm/$c$ in the case of Ref.~\cite{hasegawa2020}). Then interference among wave packets may have been lost, at least partly, and therefore the barrier passage is likely determined by independent motions of wave packets (or e.g.\ of rigid pairs or groups of wave packets) when they encounter the barrier.

The barrier passage speculated above is of classical nature, at least partly, because passage occurs mainly when the energy is higher than $V_{\text{B}}$. In View A, barrier passage occurred, simply because the initial momentum distribution $f(p)$ had included high-momentum components.  In View B, the momenta  $p_a(t)$ are changing in time during the free propagation, and the barrier passage may be determined by the value of $p_a^2(t)/2\mu$ when each wave packet arrives at the barrier. However, the interpretation has to be more complicated if wave packets still interfere around the barrier, which is beyond the scope of View B.

We can still expect that some probability of true quantum tunneling is contained in the total probability of barrier passage. It is in principle possible to investigate such possibility by analyzing the energy $E$ of the transmitted particle. In View A, due to the distribution $f(p)$, the energy may be $E>V_{\text{B}}$ in many cases but some probability may be found in $E<V_{\text{B}}$, which can be identified as quantum tunneling. In the other View B, it is expected that the energies $E_a=p_a^2/2\mu$ of the transmitted wave packets are usually higher than $V_{\text{B}}$, which however does not deny a possibility that some transmitted wave packets end up with lower energies if wave packets are still interfering around the barrier (see an example in Sec.\ \ref{sec:passage-two}). To obtain a converged result, dependence on the conditions needs careful investigation, such as the number of wave packets and the timing of arrival at the barrier.

In the approach called entangled trajectory molecular dynamics \cite{donoso2001, aswang2009, lfwang2012}, where quantum effects are taken into account by coupling the phase space trajectories, the calculated transmission probability is often compared with the classical result obtained from the same initial condition of the phase space distribution (in View A). This allows identification of the genuine quantum contribution to the barrier transmission. Similar analysis should also be performed for the TDGCM solution.

Another strict check for tunneling is to analyze the dependence on the thickness of the potential barrier $V(x)$. If the transmission is quantum tunneling, its probability should decrease when the width of the barrier is increased. The transmission is not quantum tunneling if it occurs for the infinite width [e.g.\ $V(x)=0$ for $x<0$ and $V(x)=V_{\text{B}}$ for $x>0$]. With a result of such a check, quantum tunneling would be clearly identified.

\subsection{\label{sec:passage-two} Superposition of two wave packets}

Next, let us investigate the case in which only two wave packets are superposed. We stay in View A here (see Sec.~\ref{sec:comments} for the case of View B). The behavior of the incoming state is already nontrivial as discussed in the previous section, and the cyclic motion in the incoming state is very slow in the pathological phase [from (e) through (g) to (a) in Fig.~\ref{fig:husimif}]. Therefore, the potential barrier will be encountered  with a large chance in the pathological phase. For example, if the barrier is encountered in the stage of (f), the front part of the distribution will pass over the barrier when the average momentum $\langle p\rangle$ in this part is high enough. In fact, each of the front and rear parts is similar to a Gaussian wave packet like the initial state in panel (c), and even a slightly larger $\langle p\rangle$ than the initial one may be sufficient for barrier passage (as reminded by Fig.~\ref{fig:momdst}). Such a barrier passage by a part with a frozen shape is not called tunneling in general. The same can occur in usual AMD and TDHF when $\langle p\rangle$ is high enough. This consideration on the above example seems to be consistent with the case shown in Fig.~2 of Ref.~\cite{hasegawa2020}.

After the front part passed over the potential barrier, the two parts may still continue to interfere weakly in a similar way to the evolution from (f) to (h) in Fig.~\ref{fig:husimif}. Then the transmitted part can reduce $\langle p\rangle$ resulting in an energy lower than $V_{\text{B}}$, as it should be in quantum tunneling. A question is whether the energy is guaranteed to result in a right value. In the example reported in Ref.~\cite{hasegawa2020}, this transmitted component had an energy expectation value lower than the $\nuc[4]{He}+\nuc[4]{He}$ threshold by about 2 MeV (see Fig.~4 of Ref.~\cite{hasegawa2020}), which is possible in a sense due to the attractive interaction between the nuclei. However, as the initial quantum state did not include components lower than the threshold, such an event with an inconsistent energy needs more investigation. This manifests the problem that the energy distribution, for the total state $\psi(x,t)$, is not conserved in TDGCM with the restricted form of Eq.~\eqref{eq:gcmwf}, even though the conservation of the energy expectation value is guaranteed.

\section{\label{sec:comments} Comments on Ref.~\cite{hasegawa2020}}

Our investigations in the previous sections show that quantum tunneling is not yet easy to argue in the TDGCM model, which disagrees with the quick conclusion of Ref.~\cite{hasegawa2020} that TDGCM could simulate quantum tunneling. Origins of this erroneous conclusion by Ref.~\cite{hasegawa2020} are explained below.

The important role played by the momentum distribution $f(p)$ was overlooked by Ref.~\cite{hasegawa2020}. Since the barrier height $V_{\text{B}}$ is very low compared to the width of $f(p)$ in their setup (see Fig.~\ref{fig:momdst}), the classical barrier passage is dominant, and true quantum tunneling from the narrow gray region in Fig.~\ref{fig:momdst} can only be a minor fraction. Nevertheless, the whole probability of barrier passage was misinterpreted as `quantum tunneling'.

Ref.~\cite{hasegawa2020} took View B, abandoning interpretation of the wave function $\psi(x,t)$. They chose the condition at $t=0$ so that all $p_a^2(t=0)/2\mu$ have almost the same value $E(t=0)$, and the simulation was regarded as corresponding to this beam energy. They then claimed that barrier transmission was all due to quantum tunneling because $E(t=0)<V_{\text{B}}$. This is a serious mistake made by Ref.~\cite{hasegawa2020}.  What one would claim depends on what time was regarded as the `initial' time $t_0$, because $p_a(t_0)$ depend on $t_0$ without any effect of the potential in the incoming state. It is actually impossible to simulate a fixed beam energy in TDGCM. There is no foundation for View B with $E(t=0)$ as the beam energy, when wave packets are coherently superposed. We can also expect that the result of a simulation depends on the timing of arrival at the barrier, which makes any statement questionable if it is based on only a single simulation. 

If $E(t=0)$ is still regarded as the beam energy, a strange problem is noticed in Figs.~2 and 4 of Ref.~\cite{hasegawa2020}, for the energies of `fusion' and `elastic' exit channels in a $\nuc[4]{He}+\nuc[4]{He}$ collision. The energy $E_1(t)=p_1^2(t)/2\mu$ in the `elastic' channel was much higher than the initial energy $E(t=0)$, by about 2 MeV. The same can be seen as the large slope of a trajectory in Fig.~2 of Ref.~\cite{hasegawa2020} after $t=4000$ fm/$c$. Thus the result of TDGCM is in contradiction to the fundamental fact that the energy in the elastic channel must agree with the initial one. In the `fusion' channel, the energy changed from $E(t=0)$ by about $-2$ MeV, as shown in Fig.~4 of Ref.~\cite{hasegawa2020}. Such deexcitaion would be possible if something like a $\gamma$ ray were emitted, but such an effect was not included here. Thus the energy conservation is unphysically violated, which makes it difficult to argue quantum tunneling.

\section{Summary and conclusion}

Investigations were made to evaluate the TDGCM model proposed by Ref.~\cite{hasegawa2020} as a method for many-particle quantum tunneling such as fusion of colliding nuclei, in the simplest examples equivalent to one-particle problems in one dimension. It was found that the description of the free propagation in the incoming state is nontrivial, in particular when only a few wave packets are superposed. This was visualized here by using the phase-space distribution function. Including also the ideal case of superposing many wave packets, one needs to consider the momentum width in the prepared initial wave function, to understand the behaviors of the incoming state, which then affect the passage of potential barrier. Depending on the condition of the simulation, the broad momentum distribution allows classical barrier passage by high-momentum components, which must be distinguished from true quantum tunneling.

In some case, a transmitted wave packet can end up with an energy lower than the barrier height. However, to argue quantum tunneling, the energy must be well under control by the model. In particular, consistency is required for the energies of the individual exit channels (e.g.\ of transmission and reflection). This criterion is not satisfied by the TDGCM model when a few wave packets are superposed, which should be carefully considered in practical applications. Generally, one has to analyze the energies in the incoming and outgoing states, e.g.\ to identify any contribution of true quantum tunneling in the calculated solutions of the TDGCM model. Also the dependence on the thickness of the potential barrier may be a clue to extract true information on tunneling.

Such careful analyses are left to be done. Contrary to the quick conclusion made by Ref.~\cite{hasegawa2020}, the description of many-particle quantum tunneling in time-dependent mean-field approaches still remains a challenging problem.

\section*{Acknowledgments}
The author thanks Naoto Hasegawa for discussions which confirmed the author's ideas presented in this article. This work was supported by JSPS KAKENHI Grant Numbers JP17K05432 and JP21K03528.

\bibliography{ono_nucl}
\end{document}